\documentclass[twocolumn,floatfix]{revtex4}
\usepackage{graphicx}

\def\GeV{{\rm\ GeV}}

\def\ve{\varepsilon}
\def\CG{{\cal G}}

\def\<{\langle}
\def\>{\rangle}
\def\be{\begin{equation}}
\def\ee{\end{equation}}
\def\bea{\begin{eqnarray}}
\def\eea{\end{eqnarray}}
\def\Re{\mathop{\rm Re}\nolimits}

\begin{document}
\title{A simple parameterization of two-photon exchange amplitude}
\author{Dmitry~Borisyuk}
\affiliation{Bogolyubov Institute for Theoretical Physics, 14-B Metrologicheskaya street, Kiev 03680, Ukraine}
\author{Alexander~Kobushkin}
\affiliation{Bogolyubov Institute for Theoretical Physics, 14-B Metrologicheskaya street, Kiev 03680, Ukraine}
\affiliation{National Technical University of Ukraine "Igor Sikorsky KPI", 37 Prospect Peremogy, Kiev 03056, Ukraine}
\begin{abstract}
We present a simple parameterization of the two-photon exchange amplitude
in elastic electron-proton scattering, suitable for fast numerical evaluation.
\end{abstract}
\maketitle

During last decade, the two-photon exchange (TPE) in the elastic $ep$ scattering
was actively discussed in the literature.
The TPE corrections were found to be important in various situations,
from proton radius determination to high-$Q^2$  form factor measurements.
There are also applications beyond pure $ep$ scattering:
for instance, the calculation of TPE on compound systems such as light nuclei \cite{Deuteron,Trinucleon}
requires knowledge of $ep$ and $en$ TPE amplitudes.
However, the calculation of TPE amplitudes for the elastic $ep$ scattering is technically difficult.
For example, Bernauer {\it et al.} in their recent work \cite{BernauerExp}
used low-$Q^2$ formulae from Ref.~\cite{ourLow} because they are "lending itself to an easy calculation",
though these formulae are only an approximation
and certainly are not valid at higher $Q^2$ ($> 0.1\GeV^2$ as stated in Ref.~\cite{ourLow}).
In the present note we try to eliminate this problem.
We propose a parameterization of the (calculated) TPE amplitudes,
in a simple form, suitable for quick numerical evaluation.
The parameterization has the main purpose to facilitate numerical evaluation of TPE amplitudes,
and does not necessarily have deep physical meaning.
Nevertheless, we enforce proper asymptotics at $\ve \to 1$.

Here we fit only so-called elastic contribution (which was calculated as described in Ref.~\cite{ourDisp},
using proton form factor parameterization from Ref.~\cite{ArringtonFF}).
This contribution is most well-understood and non-controversial;
moreover, it gives the dominant part of the TPE correction to the unpolarized cross-section \cite{BlundenRes}.
Though it was shown \cite{ourDelta} that inelastic contribution due to $\Delta(1232)$ intermediate state
can also give large correction to the $G_E/G_M$ ratio,
we do not consider any inelastic contributions here,
since we feel they are not estimated well enough at present time.

In any case, the knowledge of the elastic contribution is necessary for calculation
of TPE corrections to both unpolarized and polarized scattering,
and thus we hope the present work will be useful.


We use the set of TPE amplitudes $\delta\CG_E$, $\delta\CG_M$, or $\delta\CG_3$, which was introduced in Ref.~\cite{ourDisp}.
Here we consider only the real part of these amplitudes, since only real part
enters the formulae for unpolarized cross-section and double-polarization observables.
The correction to unpolarized cross-section will be
\be
 \frac{\delta\sigma}{\sigma} = 
 \frac{2}{\ve R^2 + \tau} \Re \left\{ \ve R^2 \frac{\delta\CG_E}{G_E} + \tau \frac{\delta\CG_M}{G_M} \right\}
\ee
where $\tau = Q^2/4M^2$ and $R=G_E/G_M$.
The following relations pertain to the polarization transfer experiments:
the correction to measured FF ratio $R_{exp}$,
\be
 \frac{\delta R_{exp}}{R_{exp}} = \Re \left\{ \frac{\delta\CG_E}{G_E}
    - \frac{\delta\CG_M}{G_M}
    - \frac{\ve(1-\ve)}{1+\ve} \frac{\delta\CG_3}{G_M} \right\}
\ee
and to longitudinal component of final proton polarization, $P_l$,
\be
 \frac{\delta P_l}{P_l} = - 2\ve \Re \left\{
   \frac{R^2}{\ve R^2 + \tau}
     \left( \frac{\delta\CG_E}{G_E} - \frac{\delta\CG_M}{G_M} \right)
   + \frac{\ve}{1+\ve} \frac{\delta\CG_3}{G_M}
 \right\}
\ee
More details can be found in Refs.~\cite{ourDisp, ourDelta}.

The TPE is an $O(\alpha)$ correction to the leading-order (one-photon exchange) scattering amplitude.
In turn, next-order, $O(\alpha^2)$ corrections, which we will certainly neglect,
are $O(\alpha)$ w.r.t. TPE.
Therefore the precision needed in the calculation of TPE is in any case no more than $\alpha \approx 1\%$.
Actually, our fit represents TPE amplitudes with the error less than 4\% relative or $4\cdot 10^{-4}$ absolute (whichever is larger).
Maximal deviations from the fit occur at very small and very large $Q^2$, in the intermediate $Q^2$ range the quality of fit is much better.
As an example, in Fig.1 we show TPE amplitudes along with our fit at $Q^2 = 2.5\GeV^2$.
\begin{figure}
\includegraphics[width=0.45\textwidth]{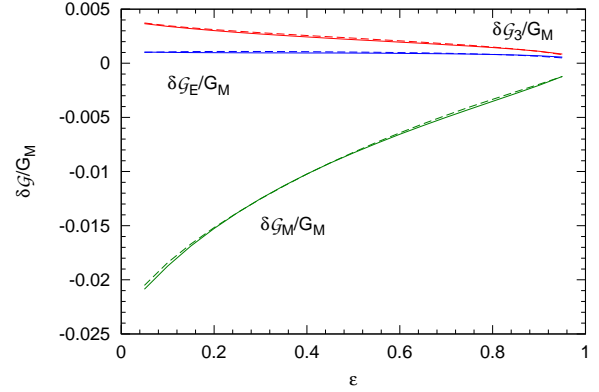}
\caption{TPE amplitudes at $Q^2=2.5\GeV^2$ (solid) and their parameterization (dashed).}
\end{figure}
\begin{table*}[!t]
\begin{tabular}{rrrrrr}
\hline\hline
\vphantom{1}\hfill$a_{i1}$\hfill\vphantom{1} &
\vphantom{1}\hfill$a_{i2}$\hfill\vphantom{1} &
\vphantom{1}\hfill$a_{i3}$\hfill\vphantom{1} &
\vphantom{1}\hfill$a_{i4}$\hfill\vphantom{1} &
\vphantom{1}\hfill$a_{i5}$\hfill\vphantom{1} &
\vphantom{1}\hfill$a_{i6}$\hfill\vphantom{1} \\
\hline\hline
  2.0126e-02 & -2.2472e-02 &  2.3423e-03 &  1.5409e-02 & -1.6884e-02 &  9.3090e-03\\
 -2.8206e-02 &  3.3562e-02 & -1.2144e-02 & -1.8363e-02 & -3.8623e-03 & -7.1703e-03\\
 -1.9260e-02 &  1.8697e-02 & -1.1005e-02 & -7.8194e-03 & -9.2961e-03 & -8.3462e-03\\
 -2.3053e-03 &  4.0685e-03 &  1.3816e-03 & -3.6264e-03 &  6.6169e-03 &  4.3448e-06\\
\hline
 -4.8590e-02 &  4.6453e-02 & -3.5080e-02 & -2.2193e-02 & -3.8962e-02 & -3.0023e-03\\
  3.1901e-02 & -2.9166e-02 &  1.4713e-02 &  1.7805e-02 &  4.5447e-03 &  6.5080e-03\\
  2.0372e-03 & -6.4414e-03 &  3.0230e-03 &  3.3810e-03 &  1.3105e-02 &  5.9247e-04\\
  3.1734e-02 & -3.3351e-02 &  1.4130e-02 &  1.5156e-02 &  3.0771e-03 &  1.5576e-03\\
\hline
  4.8470e-03 &  7.0814e-03 & -1.7742e-02 &  1.6332e-03 & -6.7499e-02 & -4.5463e-02\\
 -3.6057e-03 & -4.1289e-03 &  1.4277e-02 & -2.0706e-03 &  6.5152e-02 & -5.5648e-02\\
 -4.6275e-03 &  1.2210e-02 & -1.4221e-02 & -4.7304e-03 & -7.8823e-02 &  1.7449e-01\\
  5.0196e-03 & -2.1461e-02 &  3.1518e-02 &  5.5065e-03 &  1.3101e-01 & -1.2430e-01\\
\hline\hline
\end{tabular}
\caption{Coefficients $a_{ij}$ for $\delta\CG_E/G_M$ (top),
$\delta\CG_M/G_M$ (middle), and $\delta\CG_3/G_M$ (bottom).}\label{Tab:G}
\end{table*}


We choose to fit over the $Q^2$ range $0 < Q^2 < 8 \GeV^2$. 
At fixed $Q^2$, the $\ve$ dependence is fitted as
\be
 \delta\CG/G_M = a_1 \sqrt{1-\ve} + a_2 \sqrt{\ve(1-\ve)} + a_3 (1-\ve) + a_4 (1-\ve)^2
\ee
where $\delta\CG$ stands for $\delta\CG_E$, $\delta\CG_M$, or $\delta\CG_3$.
The function is constructed so that it vanishes at $\ve=1$, as implied by dispersion relations.
The $\sqrt{1-\ve}$ term reproduces the amplitude behaviour at $\ve \to 1$;
it is equivalent to $1/\nu$, which again follows from the dispersion integral.

The $Q^2$ dependence of the coefficients $a_i$ was parameterized as follows:
\be
 a_i = a_{i1} + a_{i2} Q^2 + (a_{i3} + a_{i4} Q^2) \ln (Q^2 + D) + 
 \frac{a_{i5}}{1+\frac{Q^2}{m^2}} + \frac{a_{i6}}{1+\frac{Q^2}{\mu^2}}
\ee
where $Q$ is in $\GeV$, $D=0.046 \GeV^2$, $m=0.359898\GeV$ and $\mu=0.095400\GeV$ were found as "best fit" values.
The coefficients $a_{ij}$ are tabulated in 
Table~\ref{Tab:G}.


\begin{thebibliography}{10}
\bibitem{Deuteron} A.P. Kobushkin, Ya.D. Krivenko-Emetov, S. Dubnicka, Phys. Rev. C {\bf 81}, 054001 (2010).
\bibitem{Trinucleon} A.P. Kobushkin, Ju. V. Timoshenko, Phys. Rev. C {\bf 88}, 044002 (2013).
\bibitem{BernauerExp}J.C.~Bernauer {\it et al.}, Phys. Rev. Lett. {\bf 107}, 119102 (2011); arXiv:1108.3533 [nucl-ex].
\bibitem{ourLow} D.~Borisyuk, A.~Kobushkin, Phys. Rev. C {\bf 75}, 038202 (2007).
\bibitem{ourDisp} D.~Borisyuk, A.~Kobushkin, Phys. Rev. C {\bf 78}, 025208 (2008).
\bibitem{ArringtonFF} J. Arrington, W. Melnitchouk, J.A. Tjon, Phys. Rev. C {\bf 76} 035205 (2007).
\bibitem{BlundenRes} S. Kondratyuk, P.G. Blunden, Phys. Rev. C {\bf 75}, 038201 (2007).
\bibitem{ourDelta} D.~Borisyuk, A.~Kobushkin, Phys.Rev. C {\bf 86}, 055204 (2012).
\end{thebibliography}
\end{document}